\documentclass[12pt,showpacs,showkeys,preprintnumbers,amsmath,amssymb,prd]{revtex4}  %,twocolumn
\usepackage{amsmath,amsfonts,amsthm,amscd,amssymb,latexsym}
\usepackage{longtable} 
\usepackage{bm}
\usepackage{dcolumn}
\usepackage{graphicx}
\usepackage{epstopdf}
\usepackage{color}
\usepackage{epsf}
\usepackage{epsfig}
\usepackage{graphicx, epic, eepic, color}

\def\beq{\begin{equation}}
\def\eeq{\end{equation}}
\def\rmd{d}
\def\rmd{{\rm d}} 

\begin{document}

\title{Peculiar velocities in dynamic spacetimes}

\author{Donato Bini}
  \affiliation{
Istituto per le Applicazioni del Calcolo ``M. Picone,'' CNR, I-00185 Rome, Italy\\
ICRA, ``Sapienza" University of Rome, I-00185 Rome, Italy
}

\author{Bahram Mashhoon}
  \affiliation{Department of Physics and Astronomy,\\
University of Missouri, Columbia, Missouri 65211, USA
}

\date{\today}

\begin{abstract}
We investigate the asymptotic behavior of peculiar velocities in certain physically significant time-dependent gravitational fields. Previous studies of the motion of free test particles have focused on the \emph{collapse scenario}, according to which a double-jet pattern with Lorentz factor $\gamma \to \infty$ develops asymptotically along the direction of complete gravitational collapse.  In the present work, we identify a second \emph{wave scenario}, in which a single-jet pattern with Lorentz factor $\gamma \to \infty$ develops asymptotically along the direction of wave propagation. The possibility of a connection between the two scenarios for the formation of cosmic jets is critically examined. 
\end{abstract}

\maketitle

\section{Introduction}
In stationary gravitational fields, a timelike Killing vector field exists such that the projection of the $4$-velocity of a free test particle on the Killing vector is a constant of the motion along the particle world line \cite{N1}. This circumstance can be interpreted to mean that there is no net exchange of energy between the particle and the gravitational field.
It is therefore a problem of basic interest whether free test particles can gain or lose energy in dynamic, i.e. time-dependent, gravitational fields. We note that the exchange of energy between charges and the electromagnetic field is a fundamental feature of electrodynamics and leads to Joule's law \cite{N2}.

To determine the energy of a test particle in the context of general relativity theory, it is necessary to refer the motion of free test particles to a set of reference observers. We take these fiducial observers to be the fundamental observers in spacetime, namely, those that are at rest in space.
We are thus interested in the peculiar velocities of free test particles relative to the class of comoving observers.

The issue of energy exchange and the nature of peculiar velocities has thus far been investigated mainly in a physical context that essentially corresponds to the modern version of the Kant-Laplace nebular hypothesis, in which the formation of elementary structure in the universe is due to the collapse of a spinning cloud of gas and dust. 
In these studies one considers exact solutions of general relativity involving certain physically significant spacetimes in which the proper distance along one spatial axis---henceforth designated
 as the $z$ axis---decreases to zero as $t\to \infty$, while the proper distances along the corresponding $x$ and $y$ axes tend asymptotically to infinity. It has been demonstrated that in such spacetimes, the timelike geodesics have a universal behavior: relative to the reference observers, free test particles asymptotically (i.e., as $t\to \infty$) form a double-jet structure along the axis of collapse and the speeds of such bulk flows tend asymptotically to the speed of light \cite{N3,N4,N5,N6}.

These {\it cosmic jets} are idealized mathematical constructs and must be clearly distinguished from astrophysical jets that are persistent high-energy magnetohydrodynamic (MHD) bipolar outflows that are generally associated with configurations that have already undergone gravitational collapse. One may hypothesize that
the collapse process is accompained by a rather mild form of the cosmic double-jet pattern, a part of which is then confined and sustained over time by various MHD mechanisms characteristic of the particular astrophysical environment.

To summarize the results of previous investigations \cite{N3,N4,N5,N6}, one may say that in a dynamic spacetime region in which asymmetric collapse/expansion is taking place, free test particles are {\it accelerated} relative to comoving observers along the collapsing direction, while they are {\it decelerated} along the expanding direction. This is in agreement with the behavior of peculiar velocities in the standard cosmological models.

It might appear that the general behavior described above is the only one that is possible in general relativity. The purpose of the present work is to show that the behavior described above is {\it not} unique. We elucidate a different  type of dynamic behavior involving a single-jet structure that is characteristic of certain propagating plane-wave spacetimes.

The plan of this paper is as follows. In Sec. II, we illustrate the nature of cosmic jets via a certain \lq\lq white-hole" interpretation of the interior Schwarzschild-Droste black hole. Sections III and IV discuss the general behavior of timelike geodesics in different plane wave spacetimes. Section V contains a discussion of our results and the possibility of a connection between these exact solutions of the gravitational field equations.

\section {Geodesics of an axially collapsing cylindrical spacetime}

Consider the standard form of the exterior Schwarzschild-Droste solution
\beq
\label{II1}
ds^2=-c^2 \left( 1-2\frac{GM}{c^2{\mathcal R}} \right)d {\mathcal T}^2+\frac{d{\mathcal R}^2}{\left( 1-2\displaystyle\frac{GM}{c^2{\mathcal R}} \right)}
+{\mathcal R}^2 (d\theta^2+\sin^2\theta d\phi^2)\,,
\eeq
where $M$ is the mass of the source. We assume that the spacetime metric has signature $+2$ and henceforth we set $c=1$. The timelike Killing vector $\partial_{\mathcal T}$ becomes null at ${\mathcal R}=2GM$ and spacelike for ${\mathcal R}<2GM$.
In this latter region of spacetime, let us introduce a new coordinate system $(t,r,\theta , \phi)$, where $t={\mathcal R}$ and $r={\mathcal T}$; moreover, we introduce a constant $T=2GM>0$. Then, metric~\eqref{II1} inside the horizon takes the form
\beq
\label{II2}
ds^2=- \frac{t}{T-t}  dt^2 +\frac{T-t}{t}dr^2+t^2 (d\theta^2+\sin^2\theta d\phi^2)\,,
\eeq
which is usually ignored in favor of the complete analytic extension of the Schwarzschild solution \cite{PK}.
Next, we introduce cylindrical coordinates $(\rho, \phi, z)$ such that
\beq
\label{II3}
\rho=L\sin \theta\,,\qquad z=r\,,
\eeq
and $\phi$ is the azimuthal angular coordinate as before.
Here $L>0$ is a constant length. Thus we express metric~\eqref{II2} as
\beq
\label{II4}
ds^2=- \frac{t}{T-t}  dt^2 +\frac{t^2}{L^2}\left( \frac{d\rho^2}{1-\rho^2/L^2}+\rho^2 d\phi^2 \right)+\frac{T-t}{t}dz^2\,.
\eeq 
This is an axially collapsing cylindrical solution of the vacuum gravitational field equations. The cylindrical axis is elementary flat and the spacetime coordinates are admissible for $0<t<T$ and $0<\rho<L$ \cite{BCM}. This solution admits two spacelike commuting Killing vector fields ($\eta=\partial_\phi$, $\zeta=\partial_z$) that are hypersurface orthogonal. There are in general four algebraic invariants of the curvature tensor for this type of Ricci-flat spacetime. In the case under study here, the only nonzero invariant is given by the Kretschmann scalar
\beq
\label{II5}
K=R_{\alpha\beta\gamma\delta}R^{\alpha\beta\gamma\delta}=12~\frac{T^2}{t^6}\,.
\eeq
Thus $t=0$ is a curvature singularity and the solution is clearly related to the Schwarzschild-Droste white hole \cite{PK}.

To study the geodesics of the cylindrical spacetime under consideration here, it proves useful to introduce standard Cartesian coordinates $(x,y,z)$, where $x=\rho \cos \phi$ and $y=\rho \sin \phi$. Thus we work with
the following metric
\beq
\label{II6}
ds^2=-\frac{t}{T-t}dt^2+\frac{t^2}{D}[(L^2-y^2)dx^2+2xydx dy+(L^2-x^2)dy^2]+\frac{T-t}{t}dz^2\,,
\eeq
where the coordinates are denoted by $x^\alpha=(t,x,y,z)$, $T$ and $L$ are two arbitrary constant parameters and 
\beq
\label{II7}
D=L^2[L^2-(x^2+y^2)]\,.
\eeq
In this case, $\sqrt{-g}=t^2/D$ and the coordinates are admissible for $0<t<T$, $|x|<L$, $|y|<L$ and $x^2+y^2<L^2$. Let us note here that the $(x, y)$ part of the metric in Eq.~\eqref{II6} can be written as $(t/L)^2 d\ell^2$, where $d\ell^2$ is the flat Euclidean 3D metric restricted to the surface of a sphere of radius $L$; see the Appendix. 
Moreover, 
 metric~\eqref{II6} is related to the Schwarzschild-Droste  {\it white hole}  
and admits two Killing vectors
\beq
\label{II8}
 \eta=-y\partial_x +x\partial_y, \qquad  \zeta=\partial_z\,,
\eeq
associated with its cylindrical symmetry.

We fix our set of fiducial observers to be at rest with respect to the spatial coordinates $(x, y, z)$. Thus these observers have 4-velocity
\beq
\label{II9}
e_{\hat 0}=\frac{1}{\sqrt{-g_{tt}}}\partial_t=\sqrt{\frac{T-t}{t}}\partial_t\,.
\eeq
The following three spatial vectors form the spatial frame of the orthonormal tetrad $e_{\hat{\alpha}}$ of our reference observers 
\begin{eqnarray}
\label{II10}
e_{\hat 1} &=& \frac{L}{t}\sqrt{\frac{L^2-x^2-y^2}{L^2-y^2}}\partial_x\,,\nonumber\\
e_{\hat 2} &=& -\frac{xy}{t\sqrt{L^2-y^2}}\partial_x +\frac{\sqrt{L^2-y^2}}{t} \partial_y\,, \nonumber\\
e_{\hat 3} &=& \sqrt{\frac{t}{T-t}}\partial_z\,.
\end{eqnarray}

Let us study the timelike geodesics of this metric, i.e. the curves with parametric equations $x^\alpha=x^\alpha (\tau)$ and with unit (timelike) tangent vector
$U^\alpha=dx^\alpha/d\tau := \dot x^\alpha$, where $\tau$ is the proper time parameter.
Decomposing $U=U^\alpha \partial_\alpha $ on the tetrad frame $e_{\hat \alpha}$, i.e., $U=U^{\hat \alpha}e_{\hat \alpha}$,  
leads to the identification of the (spatial) relative velocity  vector, $v^{\hat a}$, $a=1,2,3$, and the associated Lorentz $\gamma$ factor
\beq
\label{II11}
U=\gamma (e_{\hat{0}}+v^{\hat  a}e_{\hat  a})\,.
\eeq
Moreover, the relation between the coordinate and frame components of $U$ are 
\begin{eqnarray}
\label{II12}
U^{\hat 0}&=&\gamma=\sqrt{\frac{t}{T-t}}~\dot t\,,\nonumber\\
U^{\hat 1}&=&\gamma v^{\hat  1}=\frac{t}{\sqrt{D(L^2-y^2)}}[ (L^2-y^2)\dot x+xy \dot y]\,, \nonumber\\
U^{\hat 2}&=&\gamma v^{\hat  2}=\frac{t}{\sqrt{L^2-y^2}}\dot y\,,  \nonumber\\
U^{\hat 3}&=&\gamma v^{\hat 3}=\sqrt{\frac{T-t}{t}}~\dot z\,.
\end{eqnarray}
The Killing vectors $\eta$ and $\zeta$ ensure that  $-yU_x+xU_y$ and $U_z$ are constants of the motion, namely,
\beq
\label{II13}
\frac{t^2}{L^2}(-y\dot x+x\dot y)=C_\eta\,, \qquad \frac{T-t}{t}\dot z=C_\zeta\,.
\eeq
Therefore, one easily finds
\beq
\label{II14}
z(\tau) = z_0+C_\zeta  \int_0^\tau \frac{t(\sigma)}{T-t(\sigma)}d\sigma\,,
\eeq
and the remaining equations for geodesic motion, $U^\alpha{}_{;\mu}~U^\mu=0$, read
\begin{eqnarray}
\label{II15}
\ddot t &=& \frac{T}{2t(t-T)}(\dot t^2-C_\zeta^2)+\frac{t-T}{D}[(L^2-y^2)\dot x^2 +2xy \dot x \dot y+(L^2-x^2)\dot y^2 ], \nonumber\\
\ddot x &=& -\frac{(L^2-y^2)x\dot x^2 +(L^2-x^2)x\dot y^2}{D} -\frac{2}{t }\frac{(D\dot t+tx^2 y \dot y)\dot x}{D}\,,\nonumber\\
\ddot y &=& -\frac{(L^2-x^2)y\dot y^2 +(L^2-y^2)y\dot x^2}{D} -\frac{2}{t }\frac{(D\dot t+tx  y^2 \dot x)\dot y}{D}\,.
\end{eqnarray}
From $U^\alpha U_\alpha=-1$ and Eq.~\eqref{II14}, we get
\beq
\label{II16}
\ddot t=\frac{3T-2t}{2t(T-t)}(C_\zeta^2-\dot t^2)+\frac{T-t}{t^2}\,,
\eeq
which implies that
\beq
\label{II16a}
\dot t^2 = \frac{(t^2+C_0)(T-t)}{t^3}+C_\zeta^2\,,
\eeq
where $C_0$ is an integration constant. 

These equations imply that our fiducial observers, for example, are {\it geodesic} observers; in fact, $x=x_0$, $y=y_0$ and $z=z_0$ with $C_\eta=C_\zeta=0$ are simple solutions of Eqs.~\eqref{II14} and~\eqref{II15}. The corresponding equation for $t$, which simply follows from the normalization condition, 
can be explicitly integrated; that is, for the geodesic reference observers at rest, $C_0=0$ and
\beq
\label{II17}
\dot t=\sqrt{\frac{T-t}{t}}\,.
\eeq
Hence,
\beq
\label{II18}
\tau= \frac{\pi}{2}T-\sqrt{t(T-t)}-T\sin^{-1} \left(1-\frac{t}{T}  \right)\,,
\eeq 
so that for $t=0$, $\tau=0$, and as $t\to T$, $\tau \to \frac{\pi}{2}T$.
Moreover, the spatial frame of the fiducial observers given by Eq.~\eqref{II10} is parallel propagated along their geodesic world lines.

We now integrate Eqs.~\eqref{II15} and~\eqref{II16} numerically from $t=\epsilon T$, where $\epsilon$,  $0<\epsilon\ll 1$, is a constant, to  $t$ approaching $T$.
As time $t$ approaches $T$, the proper distance along the $z$ axis decreases to zero, while the proper distance in the $(x,y)$ plane increases in accordance with our spacetime metric~\eqref{II6}. Our numerical experiments support the general result that a double-jet structure invariably emerges along the $z$ axis with the corresponding Lorentz $\gamma$ factor approaching infinity.
That is, $(v^{\hat 1},v^{\hat 2},v^{\hat 3})\to (0,0,\pm 1)$ as $t \to T$.
The numerical results in a typical case are illustrated in Fig. 1.

\begin{figure}
\[
\begin{array}{cc}
\includegraphics[scale=0.3]{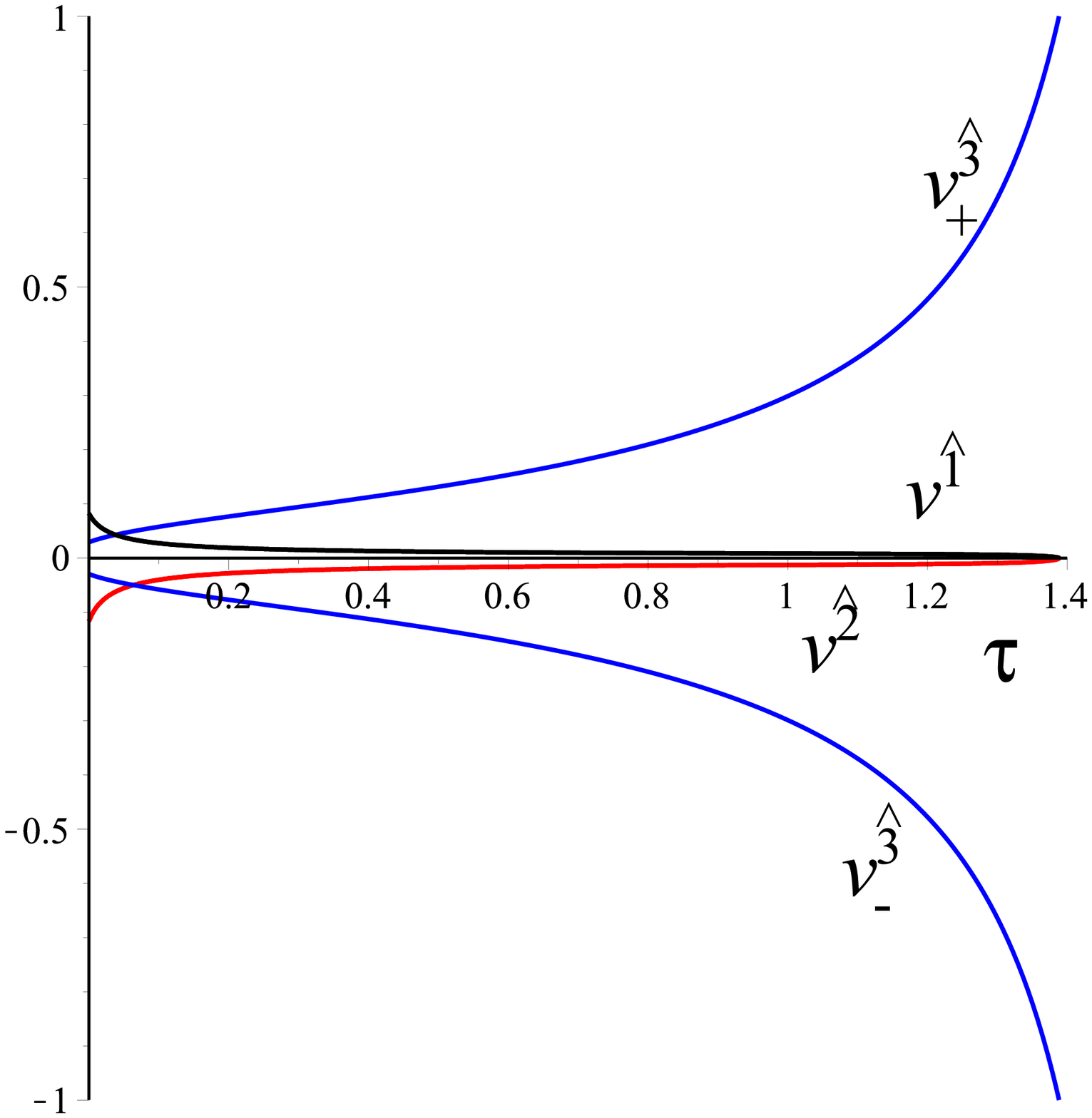} &  \qquad  \includegraphics[scale=0.3]{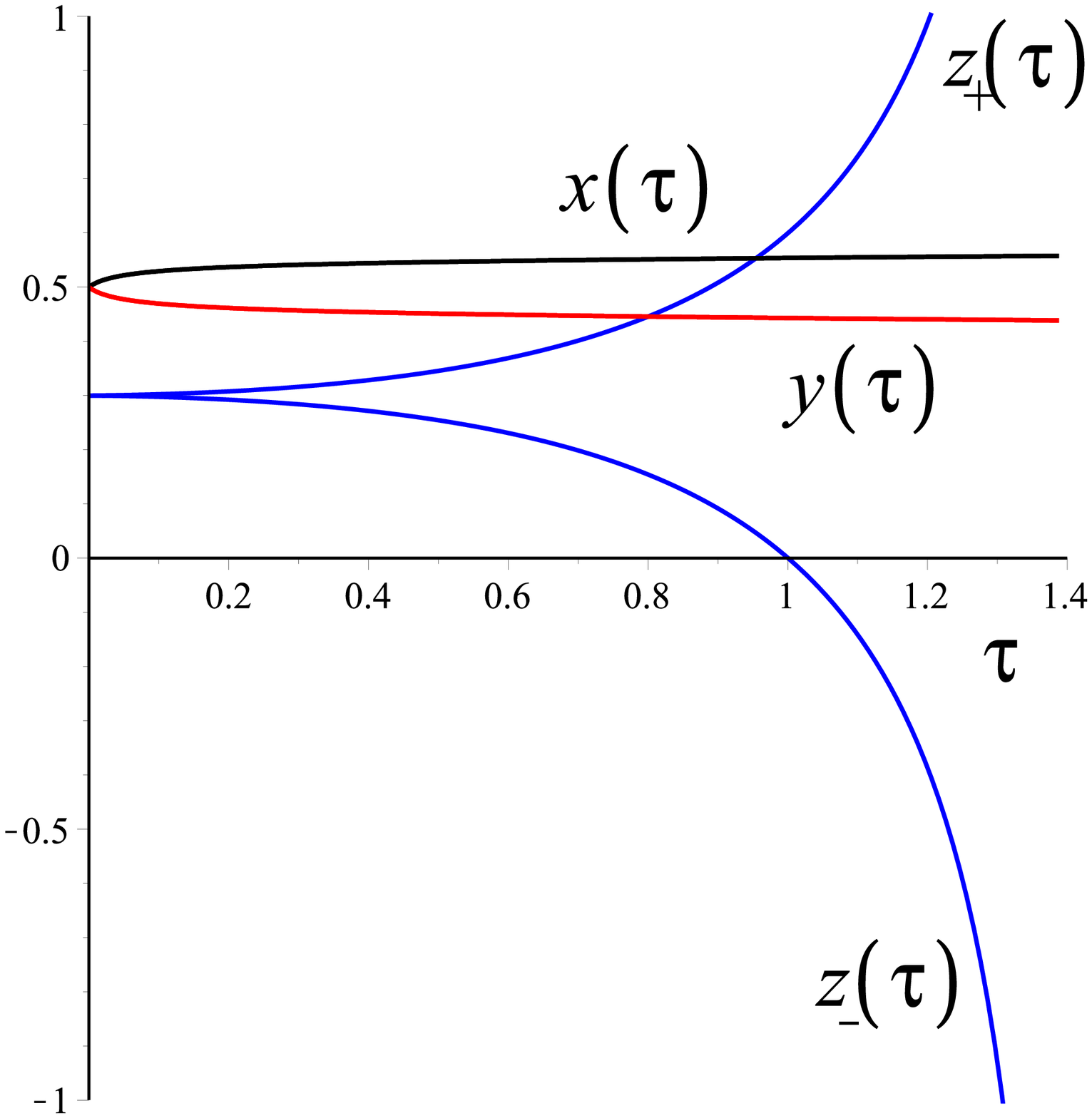} \cr
\end{array}
\]
\caption{The \emph{collapse scenario}. Left panel:  Plots of the  components of the spatial velocity $v^{\hat a}$ relative to comoving observers, given in Eq.~\eqref{II12}, versus proper time $\tau$ for the choice of metric parameters $L=T=1$.
Initial conditions at $\tau=0$ are $x(0)=y(0)=0.5$, $z(0)=0.3$, $\dot x(0)=-~\dot y(0)=1$, $\dot z_{\pm}(0)=\pm\, 0.01$ and $\dot t(0)=3$. The value of $t(0)$ is obtained from the normalization condition and is given by $t(0)=0.1019457965$.
Right panel: The initial segments of geodesic curves $x=x(\tau)$, $y=y(\tau)$ and $z=z_{\pm}(\tau)$  are plotted for the same choice of initial conditions as in the left panel.}
\end{figure}

The mathematical properties of these cosmic jets that are characteristic of the scenario for the formation of elementary astrophysical structure through gravitational collapse have been illustrated here by means of a Ricci-flat solution of general relativity that is related to the Schwarzschild-Droste white hole.
It is known that the white hole region of the extended Schwarzschild-Droste spacetime is locally isometric to colliding plane gravitational waves \cite{FI,HA}.
This circumstance provides the motivation to investigate peculiar velocities and cosmic jets in plane-wave spacetimes.

\section{Gravitational plane waves}

Consider an exact plane wave propagating along the positive $z$ direction. Introducing retarded and advanced null coordinates $u$ and $v$, respectively, by
\beq
\label{III1}
u=\frac{1}{\sqrt{2}}(t-z)\,,\qquad v=\frac{1}{\sqrt{2}}(t+z)\,,
\eeq
the spacetime metric can be expressed as
\beq
\label{III2}
ds^2=-2du dv +{\cal W}^2 (e^{2h}dx^2+e^{-2h}dy^2)\,,
\eeq
where $dt^2-dz^2=2du dv$.
Henceforth, we use the $(t,x,y,z)$ coordinate system in our discussion of plane waves. 
Here ${\cal W}(u)$ and $h(u)$ are functions of the retarded time $t-z=\sqrt{2}u$. Einstein's field equations in vacuum ($R_{\mu\nu}=0$) are satisfied provided
\beq
\label{III3}
{\cal W}_{,uu}+h_{,u}^2~{\cal W}=0\,,
\eeq
where ${\cal W}_{,u}=d{\cal W}/du$, etc. This is a Petrov type N gravitational field representing a linearly polarized (\lq\lq $\oplus$") plane wave \cite{BPR}. For $|h|\ll 1$ and ${\cal W}=1$, Eq.~\eqref{III3} is satisfied to linear order in $h$ and we recover linearized gravitational plane waves with $\oplus$ polarization in the transverse-traceless (TT) gauge.

We are interested in the peculiar velocities of timelike geodesics with respect to the class of comoving observers. It turns out that observers at rest follow timelike geodesic world lines and their natural tetrad frames, namely,
\beq
\label{III4}
e_{\hat 0}=\partial_t\,,\quad e_{\hat 1}=({\cal W}e^h)^{-1}\partial_x\,,\quad
e_{\hat 2}=({\cal W}e^{-h})^{-1}\partial_y\,,\quad e_{\hat 3}=\partial_z\,,
\eeq
are parallel propagated along their world lines.

The Riemann curvature tensor as measured by the reference observers can be expressed in any Ricci-flat spacetime in terms of $3\times 3$ symmetric and traceless matrices ${\mathcal E}$  and ${\mathcal B}$ in the standard manner as
\beq
\label{III5}
{\mathfrak R}=\left[
\begin{array}{cc}
{\mathcal E} & {\mathcal B}\cr
{\mathcal B} & -{\mathcal E}\cr
\end{array}
\right]\,,
\eeq
where ${\mathcal E}$  and ${\mathcal B}$ are the \lq\lq electric" and \lq\lq magnetic" components of the Weyl tensor. More specifically, we consider $R_{\hat\alpha\hat\beta \hat \gamma \hat \delta}$ and the mapping $(R_{\hat\alpha\hat\beta \hat \gamma \hat \delta})\mapsto (R_{IJ})$, where $I$ and $J$ range over the set $\{01,02,03,23,31,12\}$. In the case under consideration here, we find
\beq
\label{III6}
{\mathcal E}={\mathcal K}(u)\left[
\begin{array}{ccc}
0&0 & 0\cr
0&-1 & 0\cr
0&0 & 1\cr
\end{array}
\right]\,,\qquad
{\mathcal B}={\mathcal K}(u)\left[
\begin{array}{ccc}
0&0 & 0\cr
0&0 & 1\cr
0&1 & 0\cr
\end{array}
\right]\,,
\eeq
where~\cite{CM}
\beq
\label{III7}
{\mathcal K}(u)=h_{,uu}+2h_{,u}\frac{{\cal W}_{,u}}{{\cal W}}\,.
\eeq

This plane wave spacetime admits a null Killing vector $\partial_v=(\partial_t+\partial_z)/\sqrt{2}$ characteristic of the wave propagation in the $z$ direction at the speed of light and two spacelike Killing vectors $\partial_x$ and $\partial_y$ characteristic of the planar symmetry of the wave front. 
These Killing vectors are generators of an Abelian group $G_3$, which acts in  null hypersurfaces that are wave fronts given by $u=$ constant.
Moreover, the plane waves under consideration here admit two additional Killing vectors given by
\beq
\label{III8}
x\partial_v +\hat f(u) \partial_x\,,\qquad y\partial_v +\hat g(u) \partial_y\,,
\eeq
where
\beq
\label{III9}
\hat f(u)= \int^u \frac{du'}{{\mathcal F}^2(u')}\,,\qquad \hat g(u)= \int^u \frac{du'}{{\mathcal G}^2(u')}\,,
\eeq
where ${\mathcal F}={\cal W}{\rm exp}(h)$ and ${\mathcal G}={\cal W}{\rm exp}(-h)$. The five Killing vectors are generators of a group $G_5$, which has as its subgroup the Abelian group $G_3$; see the discussion of pp waves in Section 24.5 of Ref. \cite{N1}.
It follows that the projections of the $4$-velocity of free test particles $U^\alpha=dx^\alpha/d\tau$ on these Killing vectors are constants of the motion. Hence, we can write
\begin{eqnarray}
\label{III10}
&& \frac{dt}{d\tau}-\frac{dz}{d\tau}=C_v\,, \\
{}\label{III11}
&& {\cal F}^2~ \frac{dx}{d\tau}=C_x\,,\qquad {\cal G}^2~ \frac{dy}{d\tau}=C_y\,,
\end{eqnarray}
where $C_v$, $C_x$ and $C_y$ are constants.
Moreover, $U^\alpha U_\alpha=-1$ implies that $C_v \not =0$ and
\begin{eqnarray}
\label{III12}
&& \frac{dt}{d\tau}=\frac{1+C_v^2}{2C_v}+\frac{1}{2C_v}\left(\frac{C_x^2}{{\mathcal F}^2}+\frac{C_y^2}{{\mathcal G}^2}  \right) \,,\quad 
\frac{dx}{d\tau}=\frac{C_x}{{\mathcal F}^2} \,,\qquad \frac{dy}{d\tau}=\frac{C_y}{{\mathcal G}^2}\,, \\
{}\label{III13}
&&  \frac{dz}{d\tau}=\frac{1-C_v^2}{2C_v}+\frac{1}{2C_v}\left(\frac{C_x^2}{{\mathcal F}^2}+\frac{C_y^2}{{\mathcal G}^2}  \right) \,.
\end{eqnarray}
Here $u$ is a linear function of $\tau$; that is, $\sqrt{2}~u=C_v~ \tau$ + constant. For the comoving observers $C_v=1$, $C_x=C_y=0$.
The projection of $U^\alpha$ on the tetrad frame of the fiducial observers~\eqref{III4} results in
\beq
\label{III14}
U^{\hat 0}=\gamma=\frac{dt}{d\tau}\,, \quad
U^{\hat 1}=\gamma v^{\hat 1}=\frac{C_x}{\mathcal F}\,, \quad
U^{\hat 2}=\gamma v^{\hat 2}=\frac{C_y}{\mathcal G}\,, \quad
U^{\hat 3}=\gamma v^{\hat 3}=\frac{dz}{d\tau}\,.
\eeq

To proceed further, we need an explicit solution of Eq.~\eqref{III3}. To this end, let us consider~\cite{GR}
\beq
\label{III15}
{\cal W}=[\cos (\beta u)  \cosh (\beta u)]^{1/2}\,,\quad e^h=\left[ \frac{\cos (\beta u) }{\cosh (\beta u)}  \right]^{1/2}\,,
\eeq
for which Eq.~\eqref{III3} is satisfied. Positive square roots are assumed throughout in our convention. Here, $\beta$ is a constant parameter that can be chosen to be positive, $\beta >0$, with no loss in generality, and $\beta$ is related to the frequency of the gravitational wave.
Moreover, Eq.~\eqref{III15} is meaningful provided $\cos (\beta u)>0$. The curvature as measured by the reference observers is given by Eqs.~\eqref{III6} and~\eqref{III7}, where ${\mathcal K}=-\beta^2$ in this case. We note that in $(t,x,y,z)$ coordinates, $\sqrt{-g}={\cal W}^2=\cos (\beta u)  \cosh (\beta u)$. 
For the sake of definiteness, we assume $\beta u \in (-\pi/2, \pi/2)$. Let us note that as $u$ increases from, say, $u=0$ and approaches $\pi/(2\beta)$, the proper spatial distance along the $x$ direction decreases to zero, while the corresponding distance along the $y$ direction increases and $\sqrt{-g}\to 0$, since in this case
\beq
\label{III16}
{\mathcal F}(u)=\cos (\beta u)\,,\qquad {\mathcal G}(u)=\cosh (\beta u)\,.
\eeq
It is then interesting to investigate the behavior of free test particles with $C_x\not =0$ with respect to the fiducial observers.
As $u\to \pi/(2\beta)$, the asymptotic expressions for $\gamma$ and $v^{\hat a}$ are
\beq
\label{III17}
\gamma \sim \frac{C_x^2}{2C_v}\frac{1}{\cos^2(\beta u)}\,, \quad
v^{\hat 1}\sim \frac{2C_v}{C_x}\cos (\beta u)\,, \quad
v^{\hat 2}\sim \frac{2C_v C_y}{C_x^2}\frac{\cos^2(\beta u)}{\cosh (\beta u)}\,, \quad
v^{\hat 3}\sim 1\,.
\eeq
Thus the free test particles with $C_x\not=0$ in this spacetime line up asymptotically with $\gamma \to \infty$ along the direction of propagation of the wave; that is, we have a single-jet pattern with  $(v^{\hat 1},v^{\hat 2},v^{\hat 3})\to (0,0,1)$ as $u\to \pi/(2\beta)$.
It is important to remark here that the null hypersurface $u=\pi/(2\beta)$ is simply a coordinate singularity, as it occurs at the limit of admissibility of the $(t,x,y,z)$ coordinate system; nevertheless, the asymptotic jet structure has been invariantly characterized and it is therefore physically meaningful.

In a gravitational plane-wave spacetime, if the proper spatial distance tends to zero in a direction transverse to the direction of propagation of the wave, most of the free test particles in this gravitational field form a single-jet structure parallel to the direction of propagation such that the speed of the jet asymptotically approaches the speed of light. This plane-wave scenario for cosmic jet formation is entirely different from the collapse scenario. We will explore this scenario further in the next section.

\section{Electromagnetic plane-wave spacetimes}

Let us next consider the metric~\cite{BS, GR2}
\beq
\label{IV1}
ds^2 = -dt^2+ \Psi^2(u)(\rmd x^2+\rmd y^2)+dz^2\,,
\eeq
which satisfies the gravitational field equations,  $G_{\mu\nu}=8\pi G~ T_{\mu\nu}$,  with 
\beq
\label{IV2}
T_{\mu\nu}= \Phi^2(u) k_{\mu} k_{\nu}\,, \qquad k=\partial_v\,,
\eeq
where $\Phi(u)$ represents the flux of the electromagnetic radiation field and is related to $\Psi(u)$ via
\beq
\label{IV3}
\Psi_{,uu}+4 \pi G~  \Phi^2(u) \Psi =0\,.
\eeq
We note that the spacetime metric here is of the same general form as the metric of the plane wave of the previous section with $\mathcal{F}^2=\mathcal{G}^2=\Psi^2$; therefore, in addition to the isometries of the previous section, metric~\eqref{IV1} is also invariant under Euclidean rotations in the $(x, y)$ plane.  The traceless energy-momentum tensor~\eqref{IV2} can be interpreted as representing either null dust moving along the $z$ direction or a pure electromagnetic radiation field with a wave vector parallel to $k$. Adopting the latter interpretation, we assume that the potential 1-form $A^\flat$ of the null electromagnetic field is aligned with a transverse spatial direction, say the $x$ axis, so that 
\beq
\label{IV4}
A^\flat=\psi(u)\, dx\,.
\eeq
Hence, the transverse gauge condition, $\partial_\mu (\sqrt{-g}A^\mu)=0$, is satisfied and the  Faraday 2-form $F^\flat =d A^\flat $ is given by
\beq
\label{IV5}
F=\psi_{,u}(u)\, du \wedge d x\,.
\eeq
Maxwell's equations are satisfied in this case and we have
\beq
\label{IV6}
\Phi=\frac{1}{\sqrt{4 \pi}}\frac{\psi_{,u}}{\Psi}\,.
\eeq
Thus the source of the gravitational field under consideration is a linearly polarized plane electromagnetic null field $(F_{\mu\nu}F^{\mu\nu}=0)$ propagating along the $z$ direction with its electric field along the $x$ direction and its magnetic field along the $y$ direction. 

As before, it is useful to introduce a family of fiducial observers that are all at rest in space with $4$-velocity vector $e_{\hat 0}=\partial_t$. It turns out that the congruence of these fiducial observer world lines is geodesic and vorticity free, but has  nonzero expansion. Moreover,  
the natural orthonormal spatial triad adapted to such an observer is given by
\beq
\label{IV7}
e_{\hat 1}=\frac{1}{\Psi}\partial_x\,,\qquad
e_{\hat 2}=\frac{1}{\Psi}\partial_y\,, \qquad
e_{\hat 3}=\partial_z\,.
\eeq
This spatial frame is parallel propagated along the world line of the reference observer. 

The electromagnetic field, as measured by these fiducial observers, is given by the projection of the Faraday tensor onto the observers' orthonormal tetrads, namely, 
\beq
\label{IV8}
F_{\hat{\alpha}\hat{\beta}}=\frac{1}{\sqrt{2}}\frac{\psi_{,u}}{\Psi}\left[
\begin{array}{cccc}
0&1&0 & 0\cr
-1&0&0& 1\cr
0&0&0&0\cr
0&-1&0& 0\cr
\end{array}
\right]\,,
\eeq
so that the measured electric and magnetic fields are each given by $-\sqrt{2\pi} ~\Phi$ along the $x$ and $y$ axes, respectively. 

Turning now to the measurement of the gravitational field, it is clear from Eqs.~\eqref{III6} and~\eqref{III7} that the Weyl curvature tensor vanishes identically in this case and the Riemann curvature tensor is then given by
\beq
\label{IV9}
R_{\mu \nu \rho \sigma} = \frac{1}{2} (R_{\mu \rho}g_{\nu \sigma}+ R_{\nu \sigma}g_{\mu \rho}-R_{\mu \sigma}g_{\nu \rho}-R_{\nu \rho}g_{\mu \sigma})\,, 
\eeq
since the scalar curvature vanishes as well ($R=0$). In general, the measured components of the Riemann tensor can be represented as a $6\times 6$ matrix 
\beq
\label{IV10}
{\mathfrak R}=\left[
\begin{array}{cc}
{\mathcal E} & {\mathcal B}\cr
{\mathcal B^{\dagger}} & {\mathcal S}\cr
\end{array}
\right]\,,
\eeq
where $\mathcal{E}$ and $\mathcal{S}$ are symmetric $3 \times 3$ matrices and $\mathcal{B}$ is traceless. In the present case, we find that the electric and magnetic components are given by
\beq
\label{IV11}
{\mathcal E}=\kappa (u)\left[
\begin{array}{ccc}
1&0 & 0\cr
0&1 & 0\cr
0&0 & 0\cr
\end{array}
\right]\,,\qquad
{\mathcal B}=\kappa (u)\left[
\begin{array}{ccc}
0&-1 & 0\cr
1&0 & 0\cr
0&0 & 0\cr
\end{array}
\right]\,,
\eeq
while the spatial components are given by $\mathcal{S}=\mathcal{E}$. Here,
\beq
\label{IV12}
\kappa(u)=2 \pi G~ \Phi^2(u)\,.
\eeq
This gravitational field is algebraically special and of Petrov type $O$. Furthermore, as mentioned before, the Weyl tensor vanishes in this case and the metric is thus conformally flat, as can be simply verified via the transformation of the null coordinate $u$, $u \mapsto u'$, where $u'={\hat f}(u)={\hat g}(u)$ with ${\hat f}$ and ${\hat g}$ that were defined in Eq.~\eqref{III9}.

To study the general behavior of test particles in this solution of the Einstein-Maxwell equations, we need an explicit solution of Eq.~\eqref{IV3}. To this end, we consider in the rest of this section~\cite{GR2}
\beq
\label{IV13}
 \Psi=\sqrt{G}~\cos{(bu)}\,, \quad \Phi=\frac{b}{\sqrt{4 \pi G}}\,, \quad \psi=\sin{(bu)}\,, 
\eeq
where $b>0$ is a constant parameter. In this case, the measured electric and magnetic fields are each of constant magnitude $b/\sqrt{2G}$ and the measured spacetime curvature is constant as well, since $\kappa=b^2/2$. To simplify matters, we assume henceforth that $G=1$. 

The spacetime under consideration admits the following Killing vectors 
\begin{eqnarray}
\label{IV14}
\xi_{(1)}&=&\partial_v\,,\quad
\xi_{(2)}=\partial_x\,,\quad
\xi_{(3)}=\partial_y\,,\nonumber\\
\xi_{(4)}&=&-y\partial_x+x\partial_y\,,\nonumber\\	
\xi_{(5)}&=&x\partial_v+\frac{\tan{(bu)}}{b}\partial_x\,,\nonumber\\	
\xi_{(6)}&=&y\partial_v+\frac{\tan{(bu)}}{b}\partial_y\,,
\end{eqnarray}
since ${\hat f}={\hat g}=b^{-1} \tan{(bu)}$ in this case. The solution of the geodesic equations of motion is essentially the same as that given in the previous section and, repeating the same analysis as before, we recover the asymptotic single-jet pattern characteristic of the \emph{wave scenario}.  

It is interesting to examine the motion of charged particles in this Einstein-Maxwell field. Consider a test particle of inertial mass $m$ and electric charge $q$ moving in this field in accordance with the Lorentz force law 
\beq
\label{IV15}
 U^\alpha{}_{;\mu} U^\mu = {\hat q}~ F^\alpha{}_{\nu} U^\nu\,,
 \eeq
where ${\hat q}:= q/m$. In components, the equations of motion are given by
\beq
\label{IV16}
 \frac{dU^0}{d\tau} -\frac{b}{2\sqrt{2}}\sin{(2bu)}~\Sigma=-\frac{{\hat q}b}{\sqrt{2}}\cos{(bu)}~ U^1\,, \eeq
\beq
\label{IV17}
 \frac{dU^1}{d\tau} -2 b \tan{(bu)}\frac{du}{d\tau}~ U^1=-\frac{{\hat q}b}{\cos{(bu)}}~\frac{du}{d\tau}\,, 
 \eeq
\beq
\label{IV18}
 \frac{dU^2}{d\tau} -2 b \tan{(bu)}\frac{du}{d\tau}~ U^2=0\,, 
 \eeq
\beq
\label{IV19}
 \frac{dU^3}{d\tau} -\frac{b}{2\sqrt{2}}\sin{(2bu)}~\Sigma=-\frac{{\hat q}b}{\sqrt{2}}\cos{(bu)}~ U^1\,, \eeq
where $\Sigma := (U^1)^2+(U^2)^2$. It follows from Eqs.~\eqref{IV16} and~\eqref{IV19} that the quantity 
\beq
\label{IV20}
 \sqrt{2}~\frac{du}{d\tau}=U^0-U^3\, 
\eeq
is a constant of the motion $C_v$. Hence, as before, we have $\sqrt{2}~u = C_v~\tau$ + constant; moreover, it follows from $U_\alpha U^\alpha = - 1$ that $C_v\ne0$.  The above system of equations can then be easily solved and the result is 
\beq
\label{IV21}
 \frac{dt}{d\tau}=\frac{1+C_v^2}{2C_v}+\frac{1}{2C_v}\frac{\left[C_x-{\hat q}\sin{(bu)}\right]^2+C_y^2}{\cos^2{(bu)}}\,
\eeq
\beq
\label{IV22} 
\frac{dx}{d\tau}=\frac{C_x-{\hat q}\sin{(bu)}}{\cos^2{(bu)}}\,, \quad \frac{dy}{d\tau}=\frac{C_y}{\cos^2{(bu)}}\,,
\eeq
\beq
\label{IV23}
 \frac{dz}{d\tau}=\frac{1-C_v^2}{2C_v}+\frac{1}{2C_v}\frac{\left[C_x-{\hat q}\sin{(bu)}\right]^2+C_y^2}{\cos^2{(bu)}}\,.
\eeq
Thus when ${\hat q}=0$, we recover the geodesic equations of motion as would be expected from the analysis of the previous section. It is interesting to mention here that certain other physical aspects of this Einstein-Maxwell field, such as the motion of spinning test particles, have recently received attention~\cite{BG, BGHJ, BGHO}.
 
 To describe the motion of charged test particles relative to the comoving reference observers, we consider the projection of the 4-velocity $U^\mu$ on the tetrad frame $e_{\hat \alpha}$ and find
\beq
\label{IV24}
\gamma=\frac{dt}{d\tau}\,, \quad
v^{\hat 1}=\Psi \frac{dx}{dt}\,, \quad
v^{\hat 2}=\Psi \frac{dy}{dt}\,, \quad
v^{\hat 3}=\frac{dz}{dt}\,.
\eeq
Let us assume that $b u \in (-\pi/2, \pi/2)$ and note that as $u\to \pi/(2b)$, all proper lengths in the transverse direction---i.e., in the $(x, y)$ plane---tend to zero, and $\sqrt{-g}=\cos^2{(bu)}$ tends to zero as well. It follows from the inspection of Eqs.~\eqref{IV21}--\eqref{IV24} that all charged or uncharged test particles with either $C_x\ne {\hat q}$ or $C_y\ne0$ asymptotically form a single-jet structure with  $(v^{\hat 1},v^{\hat 2},v^{\hat 3})\to (0,0,1)$ as $u\to \pi/(2b)$. On the other hand, if $C_x =  {\hat q}$ and $C_y = 0$, then in this special case as $u\to \pi/(2b)$, we have $v^{\hat 1}=v^{\hat 2}= 0$ and $v^{\hat 3}=(1-C_v^2)/(1+C_v^2)$, so that $0<v^{\hat 3}<1$.  We conclude that in this \emph{wave scenario}, test particles generally line up asymptotically along the direction of wave propagation and form a single-jet pattern with Lorentz factor $\gamma \to \infty$.

\section{Concluding Remarks}

A \emph{cosmic jet} is a significant asymptotic pattern of test-particle motion in time-dependent gravitational fields such that the associated peculiar velocities generally line up in a flow with asymptotic Lorentz factor $\gamma \to \infty$.

It has been shown that cosmic double-jet patterns occur in certain exact solutions of general relativity involving gravitational collapse~\cite{N3,N4,N5,N6}. The precise relationship between these mathematical structures and the bipolar flow patterns of astrophysical jets is unknown, though in this connection certain apparently reasonable conjectures have been formulated~\cite{N4, N5, N6}.

The main purpose of this paper has been to demonstrate the existence of a second independent scenario for cosmic jet formation. That is, in addition to the \emph{collapse scenario} for the formation of \emph{double-jet} patterns~\cite{N3,N4,N5,N6}, we have presented, in the previous two sections of this paper, instances of the \emph{wave scenario}  that clearly demonstrate the generation of \emph{single-jet} patterns  in plane-wave spacetimes.  

Is there a connection between the two scenarios for cosmic jet formation? Intuitively, it is tempting to think of a time-dependent gravitational field in the collapse scenario as representing the collision of two nonlinear gravitational wave packets moving in opposite directions.  However, as discussed in the Appendix, there are severe difficulties with such a heuristic interpretation. This issue and the question of the possibility of existence of other scenarios for jet formation require further investigation.

\appendix
\section{Colliding Waves}

The illustration of the \emph{collapse scenario} in Sec. II involved a special Ricci-flat solution that is related to the Schwarzschild-Droste spacetime \emph{inside} the black hole. It has been shown in detail by Ferrari and Iba\~nez~\cite{FI} that the corresponding spacetime region is locally isometric to a colliding plane-wave spacetime; however, the plane waves in this case propagate parallel and antiparallel to a direction that is \emph{not} related to our collapse direction, which is the $z$ axis. 

In an attempt to ameliorate this situation, let us consider metric~\eqref{II6} in Sec. II and write it in the form
\beq
\label{A1}
ds^2=-\frac{t}{T-t}dt^2+\frac{T-t}{t}dz^2+ \frac{t^2}{L^2} d\ell^2,
\eeq
where
\beq
\label{A2}
d\ell^2=\frac{L^2}{D}[(L^2-y^2)dx^2+2xydx dy+(L^2-x^2)dy^2]\,.
\eeq
We note that $d\ell^2=dx^2+dy^2+dw^2$, where $w^2=L^2-x^2-y^2$; that is, $d\ell^2$ is in fact the flat Euclidean 3D metric that has been restricted to the surface of a sphere of radius $L$. Next, we define a new temporal coordinate $\Theta$ such that 
\beq
\label{A3}
\frac{dt}{d\Theta} = \frac{T-t}{t}\,.
\eeq
It follows that there is a one-to-one correspondence between $t$ and $\Theta$; that is,
\beq
\label{A4}
\Theta=-t+T\ln\left(\frac{T}{T-t}\right)\,, \qquad t=T+T~W \left(-e^{-1-\Theta/T}\right)\,,
\eeq
where we have used the principal branch of Lambert's $W$ function, which is the inverse of the function $x\mapsto xe^{x}$. Moreover,  as $t: 0\to T$, we find that $\Theta: 0 \to \infty$.
Let us note that $\Theta$ is the temporal analog of the radial ``tortoise coordinate" in the exterior  Schwarzschild-Droste spacetime. Thus in terms of $\Theta$, metric~\eqref{II6} takes the form 
\beq
\label{A5}
ds^2=\frac{T-t(\Theta)}{t(\Theta)}(-d\Theta^2+dz^2)+ \frac{t^2(\Theta)}{L^2} d\ell^2\,.
\eeq
Introducing the new null coordinates
\beq
\label{A6}
{\tilde u}=\frac{1}{\sqrt{2}}(\Theta-z)\,,\qquad {\tilde v}=\frac{1}{\sqrt{2}}(\Theta+z)\,,
\eeq
we can write the metric in the form
\beq
\label{A7}
ds^2=- 2 \chi^2 d{\tilde u}d{\tilde v}+ \Delta^2 d\ell^2\,,
\eeq
where
\beq
\label{A8}
[(T-t(\Theta))/t(\Theta)]^{1/2} := \chi({\tilde u}+{\tilde v})\,, \qquad L^{-1}t(\Theta):=\Delta({\tilde u}+{\tilde v})\,.
\eeq
Assuming that metric~\eqref{A7} can represent the collision of two nonplanar gravitational waves propagating parallel and antiparallel to the $z$ direction, it is not clear how two distinct nonlinear waves can be identified and separated out such that the wave scenario can be verified in each case~\cite{KP, UY}.

 \end{document}